\documentclass[aps,pre,twocolumn,showpacs,longbibliography,10pt,nobalancelastpage]{revtex4-1}

\usepackage{amsmath,amssymb}
\usepackage{graphicx}
\usepackage{xcolor}
\usepackage{mathrsfs}
\usepackage{newtxtext}
\usepackage[slantedGreek,cmintegrals]{newtxmath}
\usepackage{stmaryrd}
\usepackage{hyperref}
\usepackage{enumitem}
\hypersetup{colorlinks,linkcolor={blue!60!black},citecolor={blue!60!black},urlcolor={blue!60!black}}

\renewcommand{\vec}[1]{\boldsymbol{#1}}

\renewcommand{\pi}{\uppi}

\DeclareMathAlphabet{\mathcal}{OMS}{cmsy}{m}{n}
\DeclareMathAlphabet{\mathbfsfit}{\encodingdefault}{\sfdefault}{b}{it}
\DeclareMathAlphabet{\mathbfsf}{\encodingdefault}{\sfdefault}{b}{n}

\begin{document}

\title{Embryonic Inversion in \emph{Volvox carteri}: The Flipping and Peeling of Elastic Lips}
\author{Pierre A. Haas}
\email{P.A.Haas@damtp.cam.ac.uk}
\author{Raymond E. Goldstein}
\email{R.E.Goldstein@damtp.cam.ac.uk}
\affiliation{Department of Applied Mathematics and Theoretical Physics, Centre for Mathematical Sciences, \\ University of Cambridge, 
Wilberforce Road, Cambridge CB3 0WA, United Kingdom}
\date{\today}%
\begin{abstract}
The embryos of the green alga \emph{Volvox carteri} are spherical sheets of cells that turn themselves inside out at the close of their development through a programme of cell shape changes. This process of inversion is a model for morphogenetic cell sheet deformations; it starts with four lips opening up at the anterior pole of the cell sheet, flipping over and peeling back to invert the embryo. Experimental studies have revealed that inversion is arrested if some of these cell shape changes are inhibited, but the mechanical basis for these observations has remained unclear. Here, we analyse the mechanics of this inversion by deriving an averaged elastic theory for these lips and we interpret the experimental observations in terms of the mechanics and evolution of inversion.
\end{abstract}

\maketitle

\section{Introduction}
Cell sheet deformations pervade animal development~\cite{keller11}, but are constrained by local and global geometry. The local constraints are the compatibility conditions of differential geometry expressed by the Gauss--Mainardi--Codazzi equations~\cite{kreyszig}. The global constraints by contrast constitute an evolutionary freedom: by evolving different global geometries, organisms can alleviate geometric constraints. This idea is embodied in the inversion process by which the different species of \emph{Volvox} turn themselves inside out at the close of their embryonic development.

\begin{figure}[b]
\includegraphics{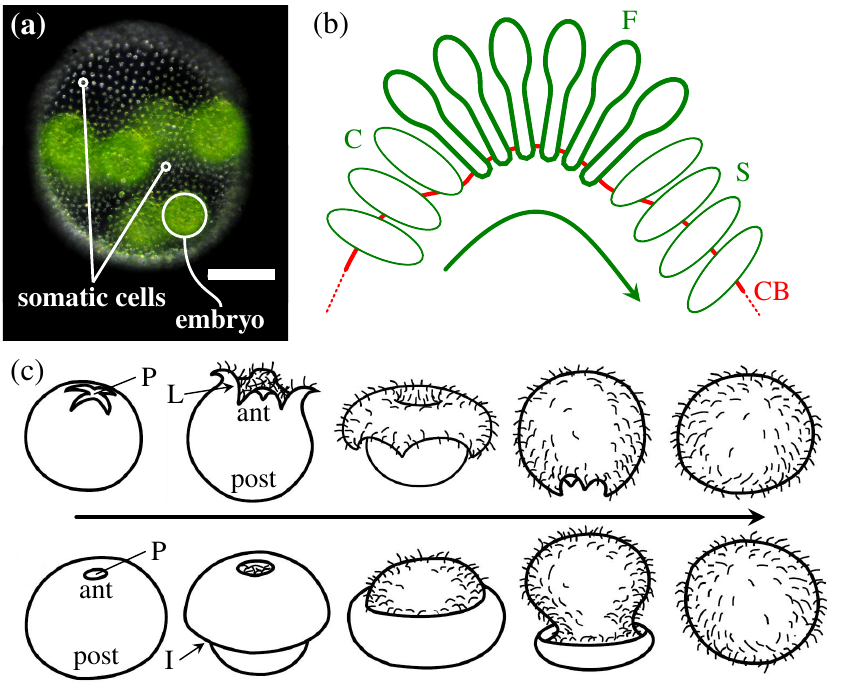}
\caption{\emph{Volvox} and its inversion. (a)~\emph{Volvox} spheroid with somatic cells and one embryo labeled. Light microscopy image by Stephanie H\"ohn reproduced from Ref.~\cite{haas15}. Scale bar: $50\,\text{\textmu m}$. (b)~Midsagittal cross-section of the cell sheet, illustrating the sequence of cell shape changes driving inversion in \emph{Volvox carteri}, following Refs.~\cite{viamontes77,viamontes79}: initially, cells are spindle-shaped~[S]. Inversion starts as a bend region of flask-shaped cells~[F] with thin stalks forms, connected by cytoplasmic bridges (CBs) at their stalks. As cells exit the bend region, they become columnar~[C]. Red line marks position of CBs; arrow marks direction of propagation of the bend region of flask-shaped cells. (c)~Inversion types, reproduced from Ref.~\cite{hohn15}. Top row: type-A inversion; bottom row: type-B inversion. Labels `ant' and `post' indicate anterior and posterior hemispheres. P: phialopore, L: lip, I: invagination. Arrow indicates the course of time.} 
\label{fig:volvox}
\end{figure}

\emph{Volvox} (Fig.~\ref{fig:volvox}a) is a genus of multicellular spherical green algae recognised as model organisms for the evolution of multicellularity~\cite{kirkbook,kirkessay,herron16,matt16} and biological fluid dynamics~\cite{goldstein15}. An adult \emph{Volvox} spheroid consists of several thousand biflagellated somatic cells and a much smaller number of germ cells or gonidia (Fig.~\ref{fig:volvox}a) embedded in an extracellular matrix~\cite{kirkbook}. The germ cells repeatedly divide to form a spherical cell sheet, with cells connected to their neighbors by the remnants of incomplete cell division, thin membrane tubes called \emph{cytoplasmic bridges}~\cite{green81a,green81b}. Those cell poles whence will emanate the flagella point into the sphere though, and so the ability to swim is only acquired once the organism turns itself inside out through a hole, the \emph{phialopore}, at the anterior pole of the cell sheet~\cite{kirkbook,hallmann06}. This process of inversion is driven by a program of cell shape changes~\cite{viamontes77,kelland77,viamontes79,hohn11}. The key cell shape change is the formation of wedge-shaped cells with thin stalks (Fig.~\ref{fig:volvox}b); at the same time, the cytoplasmic bridges move to connect the cells at their thin wedge ends~\cite{green81b}, thus splaying the cells and hence bending the cell sheet.

The precise sequence of cell sheet deformations and program of cell shape changes driving inversion varies from species to species~\cite{hallmann06}, but is broadly classified into two inversion types (Fig.~\ref{fig:volvox}c): in type-A inversion~\cite{viamontes77,viamontes79,hallmann06}, four lips open up at the anterior pole of the cell sheet, flip over, and peel back to achieve inversion. Type-B inversion~\cite{hallmann06,hohn11} starts with a circular invagination near the equator of the cell sheet, which initiates inversion of the posterior hemisphere. The phialopore then widens and the anterior hemisphere peels back over the partly inverted posterior to achieve inversion.

In spite of these differences, the mechanical crux for both inversion types is the widening of the phialopore to enable the cell sheet to pass through it. In type-A inversion, this is facilitated by the presence of the lips, and a simple program of cell shape changes suffices to invert the cell sheet~\cite{viamontes77,viamontes79}: the cells of \emph{Volvox carteri} become spindle-shaped at the beginning of inversion (Fig.~\ref{fig:volvox}b). A group of cells near the phialopore then become flask-shaped, with thin stalks. This bend region expands towards the posterior pole, leaving behind column-shaped cells (Fig.~\ref{fig:volvox}b). The program of cell shape changes in type-B inversion, by contrast, is rather more complicated, involving different types of cell shape changes in different parts of the cell sheet~\cite{hohn11}; in particular, cells and cytoplasmic bridges near the phialopore elongate in the circumferential direction to widen the phialopore. 

We have previously analyzed the mechanics of type-B inversion in detail~\cite{hohn15,haas15,haas18}, because it shares the feature of invagination with developmental events in higher organisms~\cite{keller11}, but the mechanics of type-A inversion and its lips have remained unexplored. Additionally, previous studies have revealed that type-A inversion in \emph{Volvox carteri} is arrested (a) if actomyosin-mediated contraction is inhibited chemically~\cite{nishii99} and (b) in a mutant in which the cytoplasmic bridges cannot move relative to the cells~\cite{nishii03}. The precise mechanical basis for these observations has remained unclear, however.

Here, we analyze the mechanics of the opening of the phialopore in type-A inversion by the flip-over of the lips. We derive an averaged elastic model for the lips and we relate the mechanical observations to the experimental results for \emph{Volvox carteri} referenced above. 

\section{Elastic Model}
The elastic model builds on the model we derived previously to describe type-B inversion in detail~\cite{haas18}, although the present calculation is more intricate because axisymmetry is broken owing to the presence of the lips. We consider a spherical shell of radius $R$ and uniform thickness $h\ll R$ (Fig.~\ref{fig:geometry}a), characterised by its arclength $s$ and distance from the axis of revolution $\rho(s)$. Cuts in planes containing the axis of symmetry divide part of the shell into $N$ lips of angular extent $2\varphi=2\pi/N$ (Fig.~\ref{fig:geometry}b). 
\subsection{The Differential Geometry of Lips}
\begin{figure}
\centering\includegraphics{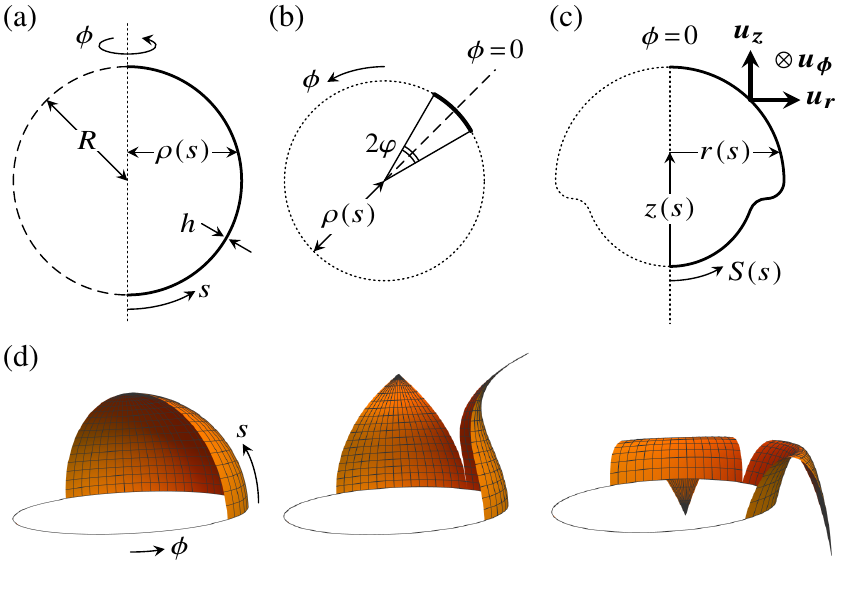}
\caption{Elastic Model. (a) Undeformed geometry: a spherical shell of radius $R$ and thickness $h\ll R$ is characterized by its arclength $s$ and distance from the axis of revolution $\rho(s)$. (b)~Anterior view of lips: cuts in planes containing the axis of symmetry define $N$ lips extending over $-\varphi\leqslant\phi\leqslant\varphi$. (c)~Deformed configuration: the midplane $\phi=0$ of the lip is characterized by its arclength $S(s)$ and distance $r(s)$ from the axis of revolution. A local basis $(\vec{u_r},\vec{u_\phi},\vec{u_z})$ describes the deformed surface. (d)~Deformation of two lips under the geometric simplification (\ref{eq:phibar}). The circumferential curvature changes sign at the point where the lip is perpendicular to the axis of revolution.}
\label{fig:geometry}
\end{figure}

We start by considering a single lip, $-\varphi\leqslant\phi\leqslant\varphi$, where $\phi$ is the azimuthal angle of the undeformed sphere. Compared to an azimuthally complete shell, the cuts allow an additional deformation mode of the shell, one of azimuthal compression or expansion. Here, we restrict to the simple deformation of \emph{uniform} stretching or compression, so that the azimuthal angle in the deformed configuration of the shell is
\begin{align}
\overline{\phi}=\Phi(s)\phi. \label{eq:phibar}
\end{align}
In the deformed configuration, the distance from the axis of revolution is $r(s)$, and the coordinate along the axis is $z(s)$ (Fig.~\ref{fig:geometry}c). These are assumed to be independent of $\phi$. This is a geometric simplification that nonetheless ensures coupling of the meridional and circumferential deformations; we discuss the basis for this approximation in more detail in Appendix~\ref{appB}. With this simplification, points on the lip initially at the same distance from and along the axis of revolution remain at the same distance from and along the axis of revolution as the lip deforms (Fig.~\ref{fig:geometry}d). 

The deformed arclength, however, is a function of both $s$ and $\phi$, and we define $S(s)$ to be the arclength of the midline $\phi=0$ of the lip. The meridional and circumferential stretches of the midline of the lip are
\begin{align}
&f_s(s)=\dfrac{\mathrm{d}S}{\mathrm{d}s},&&f_\phi(s)=\dfrac{r(s)}{\rho(s)}. 
\end{align}

\noindent The position vector of a point on the midsurface of the deformed shell is thus
\begin{equation}
\vec{r}(s,\phi)=r(s)\vec{u_r}\bigl(\Phi(s)\phi\bigr)+z(s)\vec{u_z},\label{eq:surf} 
\end{equation}
in a right-handed set of axes $(\vec{u_r},\vec{u_\phi},\vec{u_z})$ and so the tangent vectors to the deformed midsurface are
\begin{align}
\vec{e_s}&=r'\vec{u_r}+r\Phi'\phi\,\vec{u_\phi}+z'\vec{u_z},&&\vec{e_\phi}=r\Phi\,\vec{u_\phi},
\end{align}
where dashes denote differentiation with respect to $s$. By definition, $r'^2+z'^2=f_s^2$, and so we may write 
\begin{align}
&r'=f_s\cos{\beta},&& z'=f_s\sin{\beta}.
\end{align}
The metric of the midsurface is thus
\begin{align}
\Bigl\{f_s^2+r^2\Phi'^2\phi^2\Bigr\}\,\mathrm{d}s^2+r^2\Phi^2\,\mathrm{d}\phi^2+2r^2\Phi\Phi'\phi\,\mathrm{d}s\,\mathrm{d}\phi,
\end{align}
and its second fundamental form is
\begin{align}
&\left\{f_s\beta'+r{\Phi'}^2\phi^2\sin{\beta}\right\}\,\mathrm{d}s^2+r\Phi^2\sin{\beta}\,\mathrm{d}\phi^2\nonumber\\
&\hspace{45mm}+2r\Phi\Phi'\phi\sin{\beta}\,\mathrm{d}s\,\mathrm{d}\phi. 
\end{align}
Further, the unit normal to the deformed midsurface is
\begin{align}
\vec{n}=\cos{\beta}\,\vec{u_z}-\sin{\beta}\,\vec{u_r}.
\end{align}
The Weingarten relations~\cite{kreyszig} yield
\begin{align}
&\dfrac{\partial\vec{n}}{\partial s}=-\kappa_s\vec{e_s}+\dfrac{\Phi'\phi}{\Phi}(\kappa_s-\kappa_\phi)\vec{e_\phi},&&\dfrac{\partial\vec{n}}{\partial\phi}=-\kappa_\phi\vec{e_\phi},
\end{align}
wherein
\begin{align}
&\kappa_s=\dfrac{\beta'}{f_s},&& \kappa_\phi=\dfrac{\sin{\beta}}{r}
\end{align}
are the principal curvatures of the midline of the lip. Hence the principal curvatures of the lip are those of its midline, though the directions of principal curvature no longer coincide with $\vec{e_s}$ and $\vec{e_\phi}$ away from the midplane $\phi=0$ of the lips.

\subsection{Calculation of the Elastic Strains}
To calculate the deformation gradient, we make the Kirchhoff `hypothesis'~\cite{audolypomeau}, that the normals to the undeformed midsurface remain normal to the midsurface in the deformed configuration of the shell. Taking a coordinate $\zeta$ across the thickness of the shell, the position vector of a general point in the shell is
\begin{align}
\vec{r}(s,\phi,\zeta)&=r\vec{u_r}+z\vec{u_z}+\zeta\vec{n}.
\end{align}
The tangent vectors to the shell are thus
\begin{subequations}
\begin{align}
\vec{e_s}&= f_s(1-\kappa_s\zeta)(\cos{\beta}\,\vec{u_r}+\sin{\beta}\,\vec{u_z})\nonumber\\
&\hspace{30mm}+\rho \Phi'\phi f_\phi(1-\kappa_\phi\zeta)\vec{u_\phi},\\
\vec{e_\phi}&= \rho\Phi f_\phi(1-\kappa_\phi\zeta)\vec{u_\phi}.
\end{align}
\end{subequations}
The geometric deformation gradient tensor is therefore
\begin{align}
\mathbfsfit{F^g}=\left(\begin{array}{cc}
f_s\bigl(1-\kappa_s\zeta\bigr)&0\\
\rho\Phi'\phi f_\phi\bigl(1-\kappa_\phi\zeta\bigr)&\Phi f_\phi(1-\kappa_\phi\zeta\bigr)
\end{array}\right).
\end{align}
Cell shape changes impart intrinsic stretches $\smash{f_s^0,f_\phi^0}$ and curvatures $\smash{\kappa_s^0,\kappa_\phi^0}$ to the cell sheet that are different from its undeformed stretches and curvatures, but the cell shape changes do  not lead to any intrinsic azimuthal compression. Hence we define the intrinsic deformation gradient tensor
\begin{align}
\mathbfsfit{F}^{\mathbfsf{0}}=\left(\begin{array}{cc}
f_s^0\bigl(1-\kappa_s^0\zeta)&0\\
0&f_\phi^0\bigl(1-\kappa_\phi^0\zeta\bigr)
\end{array}\right),
\end{align}
Invoking the standard multiplicative decomposition of morphoelasticity~\cite{goriely}, the elastic deformation gradient tensor $\mathbfsfit{F}$ is defined by $\mathbfsfit{F^g}=\mathbfsfit{F}\mathbfsfit{F}^{\mathbfsf{0}}$. Hence the Cauchy--Green tensor is
\begin{widetext}
\begin{align}
\mathbfsfit{C}=\mathbfsfit{F}^\top\mathbfsfit{F}=\left(\begin{array}{cc}\vspace{1mm}
\left(\dfrac{f_s(1-\kappa_s\zeta)}{f_s^0(1-\kappa_s^0\zeta)}\right)^2+\rho^2{\Phi'}^2\phi^2\left(\dfrac{f_\phi(1-\kappa_\phi\zeta)}{f_s^0(1-\kappa_s^0\zeta)}\right)^2&\dfrac{\rho\Phi'\Phi\phi f_\phi^2(1-\kappa_\phi\zeta)^2}{f_s^0f_\phi^0(1-\kappa_s^0\zeta)(1-\kappa_\phi^0\zeta)}\\
\dfrac{\rho\Phi'\Phi\phi f_\phi^2(1-\kappa_\phi\zeta)^2}{f_s^0f_\phi^0(1-\kappa_s^0\zeta)(1-\kappa_\phi^0\zeta)}&\left(\dfrac{\Phi f_\phi(1-\kappa_\phi\zeta)}{f_\phi^0(1-\kappa_\phi^0\zeta)}\right)^2
\end{array}\right), 
\end{align}
from which we derive the elastic strains $\smash{\vec{\varepsilon}=\tfrac{1}{2}(\mathbfsfit{C}-\mathbfsfit{I})}$. While we do not make any assumptions about the \emph{geometric} or \emph{intrinsic} strains associated with $\mathbfsfit{F^g}$ or $\mathbfsfit{F}^{\mathbfsf{0}}$, respectively, we assume that these \emph{elastic} strains remain small. We therefore approximate the strains along the midline by
\begin{align}
f_s(1-\kappa_s\zeta) &\approx f_s^0(1-\kappa_s^0\zeta),&f_\phi(1-\kappa_\phi\zeta)&\approx f_\phi^0(1-\kappa_\phi^0\zeta)/\Phi,
\end{align}
except in differences of these expressions, so that
\begin{align}
\varepsilon_{ss}&\approx \dfrac{f_s(1-\kappa_s\zeta)}{f_s^0(1-\kappa_s^0\zeta)}-1 + \tfrac{1}{2}\rho^2{\Psi'}^2\phi^2\left(\dfrac{f_\phi^0(1-\kappa_\phi^0\zeta)}{f_s^0(1-\kappa_s^0\zeta)}\right)^2,&\varepsilon_{\phi\phi}&\approx\dfrac{\Phi f_\phi(1-\kappa_\phi\zeta)}{f_\phi^0(1-\kappa_\phi^0\zeta)}-1,&\varepsilon_{s\phi}&\approx\varepsilon_{\phi s}\approx \tfrac{1}{2}\rho\Psi'\phi\dfrac{f_\phi^0(1-\kappa_\phi^0\zeta)}{f_s^0(1-\kappa_s^0\zeta)},
\end{align}
where we have introduced $\Psi=\log{\Phi}$.
\subsection{Calculation of the Elastic Energy}
To derive the elastic energy, we need to specify the constitutive relations. As in our previous work~\cite{hohn15,haas15,haas18}, we assume that the shell is made of a Hookean material~\cite{libai,audolypomeau}, characterized by its constant elastic modulus $E$ and its Poisson ratio~$\nu$. The elastic modulus appears only as an overall constant that ensures that $\mathcal{E}$ has units of energy. We shall assume moreover that $\nu=1/2$ for incompressible biological material. The elastic energy (per unit extent in the meridional direction) is thus
\begin{align}
\dfrac{\mathcal{E}}{2\pi\rho}&=\dfrac{E}{2(1-\nu^2)}\int_{-h/2}^{h/2}{\left\{\vphantom{\left(\dfrac{A^2}{A^2}\right)^2}\dfrac{1}{2\varphi}\int_{-\varphi}^{\varphi}{\Bigl[(\varepsilon_{ss}+\varepsilon_{\phi\phi})^2-2(1-\nu)\bigl(\varepsilon_{ss}\varepsilon_{\phi\phi}-\varepsilon_{s\phi}\varepsilon_{\phi s}\bigr)\Bigr]\,\mathrm{d}\phi}\right\}\mathrm{d}\zeta}. 
\end{align}
Performing the integrals across $\zeta$ and $\phi$ and expanding up to and including third order in $h$, we obtain
\begin{align}
\dfrac{\mathcal{E}}{2\pi\rho}&=\dfrac{Eh}{2(1-\nu^2)}\Biggl\{E_s^2+E_\phi^2+2\nu E_sE_\phi+2a\lambda^2{\Psi'}^2\bigl(E_s+\nu E_\phi\bigr)+a(1-\nu)\lambda^2{\Psi'}^2+\tfrac{9}{5}a^2\lambda^4{\Psi'}^4\Biggr\}\nonumber\\
&\hspace{5mm}+\dfrac{Eh^3}{24(1-\nu^2)}\Biggl\{K_s^2+K_\phi^2+2\nu K_sK_\phi+3{\kappa_s^0}^2E_s^2+3{\kappa_\phi^0}^2E_\phi^2+2\nu\Bigl({\kappa_s^0}^2+\kappa_s^0\kappa_\phi^0+{\kappa_\phi^0}^2\Bigr)E_sE_\phi-4\kappa_s^0 E_sK_s\nonumber\\
&\hspace{30mm}-4\kappa_\phi^0 E_\phi K_\phi-2\nu\bigl(\kappa_s^0+\kappa_\phi^0\bigr)\bigl(E_sK_\phi+E_\phi K_s\bigr)+2a\lambda^2{\Psi'}^2\biggl[\Bigl(6{\kappa_s^0}^2-6\kappa_s^0\kappa_\phi^0+{\kappa_\phi^0}^2\Bigr)E_s\nonumber\\
&\hspace{70mm}+\nu \kappa_s^0\bigl(3\kappa_s^0-2\kappa_\phi^0\bigr)E_\phi-\bigl(3\kappa_s^0-2\kappa_\phi^0\bigr)K_s-\nu\bigl(2\kappa_s^0-\kappa_\phi^0\bigr)K_\phi\biggr]\nonumber\\
&\hspace{45mm}+a(1-\nu)\lambda^2{\Psi'}^2\Bigl(3{\kappa_s^0}^2-4\kappa_s^0\kappa_\phi^0+{\kappa_\phi^0}^2\Bigr)+\tfrac{9}{5}a^2\lambda^4{\Psi'}^4\Bigl(10{\kappa_s^0}^2-16\kappa_s^0\kappa_\phi^0+6{\kappa_\phi^0}^2\Bigr)\Biggr\},\label{eq:E}
\end{align}
wherein $a=\pi^2\big/6N^2$ and $\lambda=\rho f_\phi^0/f_s^0$, and the shell strains and curvature strains are defined by
\begin{align}
E_s&=\dfrac{f_s-f_s^0}{f_s^0},&K_s&=\dfrac{f_s\kappa_s-f_s^0\kappa_s^0}{f_s^0},&E_\phi&=\dfrac{\Phi f_\phi-f_\phi^0}{f_\phi^0},&K_\phi&=\dfrac{\Phi f_\phi\kappa_\phi-f_\phi^0\kappa_\phi^0}{f_\phi^0}.
\end{align}
\end{widetext}
We derive the governing equations associated with this energy in appendix~\ref{appA}. We solve these equations numerically using the boundary-value problem solver \texttt{bvp4c} of \textsc{Matlab} (TheMathworks, Inc.) and the continuation software \textsc{Auto}~\cite{auto}.
\section{Results}

\begin{table}[b]
\caption{\label{tab}Measurements of geometric parameters for type-A inversion in \emph{Volvox carteri} from previous measurements or extracted from previously published experimental figures. BR: bend region.}
\begin{ruledtabular}
\begin{tabular}{lll}
Quantity & Measurement & Reference\\
\hline
Number of lips $N$ & $4$ & \cite{hallmann06}; \cite{viamontes79}, Fig.~2c \\
Radius $R$\footnote{Radius of cell sheet at phialopore opening.} & $\sim 35\,\text{\textmu m}$&\cite{hallmann06}, Fig.~8b; \cite{viamontes79}, Fig.~2f\\
Thickness $h$\footnote{Thickness of cell sheet at phialopore opening.} & $\sim 7\,\text{\textmu m}$ &\cite{viamontes79}, Fig.~2f, Tab.~1\\
Cell widths\footnote{Cell widths are estimated along the midline of the cell sheet.} & & \\
\phantom{.}\hspace{3mm}spindle cells & $2.14\,\text{\textmu m}$& \cite{viamontes79}, Table 1\\
\phantom{.}\hspace{3mm}flask cells & $2.05\,\text{\textmu m}$& calc. from \cite{viamontes79}, Table 1\\
\phantom{.}\hspace{3mm}columnar cells & $2.36\,\text{\textmu m}$& \cite{viamontes79}, Table 1\\
Phialopore $P$ & $\sim 0.3$ & \cite{viamontes79}, Fig.~2f; \cite{nishii99}, Fig.~1b\\
Fraction of cells in BR & $\sim 0.25$ &\cite{viamontes77}, Fig.~2c--e
\end{tabular}
\end{ruledtabular}
\end{table}
\subsection{Estimates of Model Parameters}
We now specialize our model to describe type-A inversion in \emph{Volvox carteri} by, in particular, encoding the observed cell shape changes (Fig.~\ref{fig:volvox}b) into the functional forms of the intrinsic stretches and curvatures. We shall introduce a number of parameters for this purpose; we base our estimates of these parameters on the measurements in Table~\ref{tab}. Some of the values in Table~\ref{tab} are taken from the literature, others are extracted from figures in the literature. In particular, these extracted values should not be taken as estimates of average values, but rather as indications of which values can be realized experimentally.

\begin{figure}[b]
\includegraphics{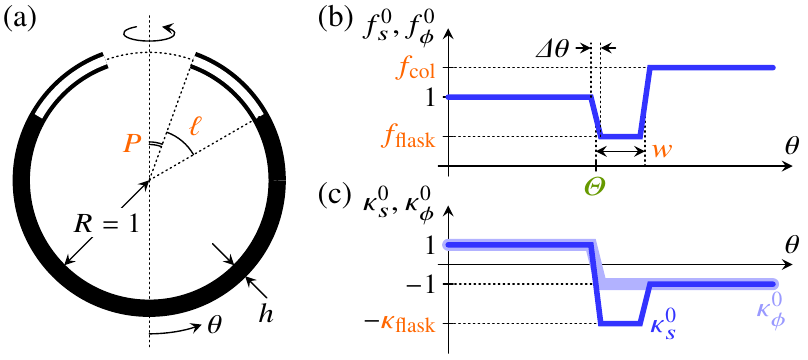}
\caption{Definition of Model Parameters (a) Undeformed configuration after formation of spindle-shaped cells: the phialopore has opened by an angle $P$ on a spherical shell of radius $R$ and thickness $h$. The lips span an angle $\ell$. (b)~Functional forms of the intrinsic stretches $\smash{f_s^0,f_\phi^0}$, as functions of the polar angle $\theta$. Parameters $f_{\text{flask}}$ and $f_{\text{col}}$ define the intrinsic stretches of the flask- and column-shaped cells relative to the spindle-shaped cells. The width of the bend region is $w$, and the posterior limit of the bend region is at $\theta=\Theta$. (c)~Functional forms of the intrinsic curvatures $\smash{\kappa_s^0,\kappa_\phi^0}$, as functions of the polar angle $\theta$ and in units where $R=1$. The intrinsic meridional curvature in the bend region is $-\kappa_{\text{flask}}$. For numerical convenience, discontinuities in the intrinsic stretches and curvatures are regularized over a small angular extent $\Delta\theta=0.1$.}
\label{fig:model}
\end{figure}

\begin{figure*}
\centering\includegraphics{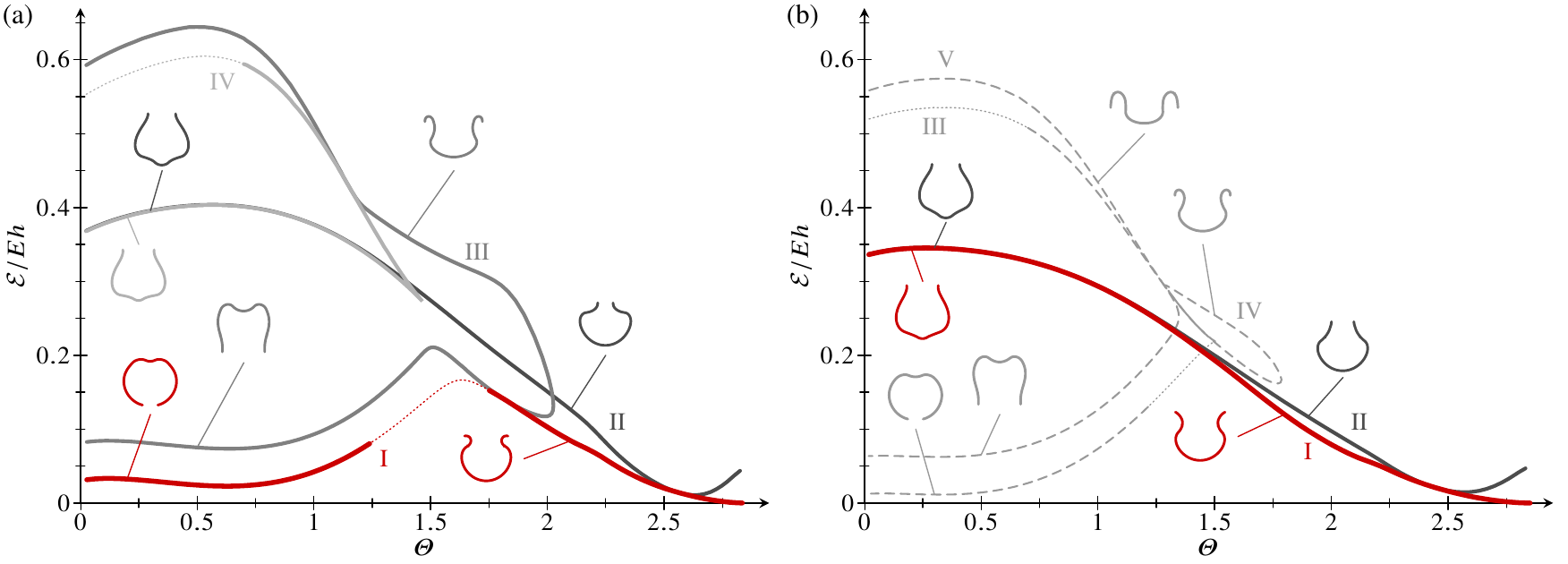}
\caption{Bifurcation behavior of the lips. Diagrams in $(\Theta,\mathcal{E})$ space of some solution branches for (a) $k=4.0>k_\ast$ and (b) $k=2.5<k_\ast$. Roman numerals label branches. The thickest, red line (branch I) corresponds to the branch of lowest energy connected to the initial state of the shell. Dashed lines in (b) indicate branches that are not connected to the lowest-energy branch. Insets show some solution shapes. On dotted parts of branches, solution shapes self-intersect.}
\label{fig:bif} 
\end{figure*}

To describe the geometry of the undeformed shell (Fig.~\ref{fig:model}a), we must specify the number $N$ of lips, the relative thickness $h/R$ of the cell sheet, the opening angle $P$ of the phialopore, and the extent $\ell$ of the lips. In accordance with the measurements in Table~\ref{tab}, we choose
\begin{align}
&N = 4,&&h/R = 0.21, && P = 0.3, && \ell = 0.7.
\end{align}
The estimate of $\ell$ is not based on a measured value, since it is rather hard to visualize the precise extent of the lips, but it is in qualitative agreement with experimental visualisations of the lips~\cite{green81a,hallmann06,viamontes77}. Lips are clearly visible before inversion starts~\cite{green81a}, but additional breaking of cytoplasmic bridges could increase $\ell$ during inversion. While breaking of cytoplasmic bridges was suggested as a possible mechanism to explain the cell rearrangements observed near the phialopore in type-B inversion~\cite{haas18}, what experimental data there are~\cite{green81a,hallmann06,viamontes77} suggest that this effect is at most small in type-A inversion, justifying the absence of a `fracture criterion' for cytoplasmic bridges in the model. 

The remaining parameters describe the functional forms of the intrinsic stretches and curvatures of the shell~(Fig.~\ref{fig:model}b,c): from measurements of the cell widths (Table~\ref{tab}), we estimate the stretches $f_{\text{flask}}$ and $f_{\text{col}}$ corresponding to flask and columnar cells (relative to spindle-shaped cells). The width $w$ of the bend region can be estimated from the fraction of flask-shaped cells in a mid-sagittal cross-section of the cell sheet (Table~\ref{tab}). Our estimates for these parameters are therefore
\begin{align}
&f_{\text{flask}}=0.95, &&f_{\text{col}}=1.1,&&w=0.65.
\end{align}
Note that we may only read the actual stretches, as opposed to the intrinsic stretches, off the deformed shapes, but since stretching is energetically more costly than bending, we expect the approximations involved in obtaining these parameter estimates from cell size measurements to be good.

We do not estimate the final parameter, $\kappa_{\text{flask}}$, the intrinsic meridional curvature of the flask cells, which is the main parameter that we vary in the analysis that follows.

\subsection{To flip, or not to flip, ...}
We fix the value of $k=\kappa_{\text{flask}}$, and propagate the bend region from the tip of the lips to their base and then towards the posterior pole by decreasing the value of parameter $\Theta$ (Fig.~\ref{fig:model}b,c) that describes the position of the bend region, starting from a nearly undeformed shell. Solution branches in $(\Theta,\mathcal{E})$ space are shown in Fig.~\ref{fig:bif}; there is a critical value $k_\ast$ separating two kinds of behavior. If $k>k_\ast$, the shell inverts on the branch of lowest energy (Branch I in Fig.~\ref{fig:bif}a). Several branches bifurcate off the latter (Branches II--IV in Fig.~\ref{fig:bif}a), but these have higher energy. If $k<k_\ast$, the shell does not invert on the branch of lowest energy or the branch connected to it (Branches I,II in Fig~\ref{fig:bif}b). There do exist branches on which the shell inverts (Branches III--V in Fig.~\ref{fig:bif}b) analogous to those in Fig.~\ref{fig:bif}a, but these are not connected to the initial state of the shell. The topology of these additional branches undergoes another bifurcation, not discussed here, as $k$ is reduced further. 

Some solution shapes on the branches in Fig.~\ref{fig:bif} self-intersect; we expect the corresponding parts of the branches to be replaced with configurations of the shell where the rim of the lips is in contact with the uninverted part of the cell sheet. In these configurations, axisymmetry is necessarily broken in the uninverted part of the shell; we do not pursue this further, although we note that we have previously analyzed an analogous contact problem in the absence of lips~\cite{haas18}. These configurations will not in fact be important for the discussion that follows. Finally, we note that no such self-intersecting configurations arise on Branch III in Fig.~\ref{fig:bif}a, the solutions on which do lead to a completely inverted shell.

\begin{figure}
\centering\includegraphics{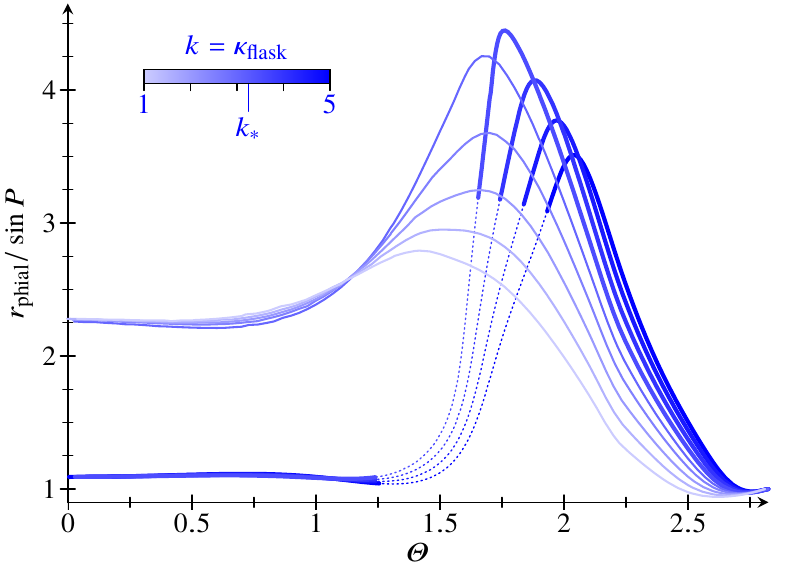}
\caption{Dynamics of lips. Radius of the phialopore $r_{\text{phial}}$ (normalized with initial phialopore radius $\sin{P}$) against $\Theta$ on Branch I, for different values of $k=\kappa_{\text{flask}}$. On branches with  $k>k_\ast$ (thick lines), self-intersecting solution shapes (dotted lines) arise. No such self-intersections arise on branches with $k<k_\ast$ (thin lines).}
\label{fig:rphial}
\end{figure}

The dynamics of the flip over of the lips on Branch I are illustrated in Fig.~\ref{fig:rphial}: as $k$ is reduced towards $k_\ast$, the lips open wider and wider before they flip over; after flip over, the opening of the phialopore decreases quickly. As $k$ is reduced below $k_\ast$, the maximal opening of the lips decreases; they do not flip over and the phialopore remains wide open.

Nishii \emph{et al.}~\cite{nishii03} showed that the InvA mutant of \emph{Volvox carteri} fails to invert. In this mutant, there is no relative motion between cells and cytoplasmic bridges, and so the flask shaped cells are not connected at their thin tips only~\cite{nishii03}. Thus the splay imparted, in the wild-type, by the combination of cell shape change and motion of cytoplasmic bridges is reduced. This corresponds, in our model, to the intrinsic curvature $\kappa_{\text{flask}}$ being reduced in the InvA mutant. The mechanical bifurcation discussed above can thus rationalize the failure of the mutant to invert. The sequence of shapes on Branch I of Fig.~\ref{fig:bif}b is indeed in excellent qualitative agreement with that observed during `inversion' of the InvA mutant, shown in Fig.~1f of Ref.~\cite{nishii03}: the lips begin to curl over, but as `inversion' progresses, the lips do not flip over and the phialopore remains wide open at the end of inversion.

\subsection{The Importance of Being Contracted}
\begin{figure}[b]
\centering\includegraphics{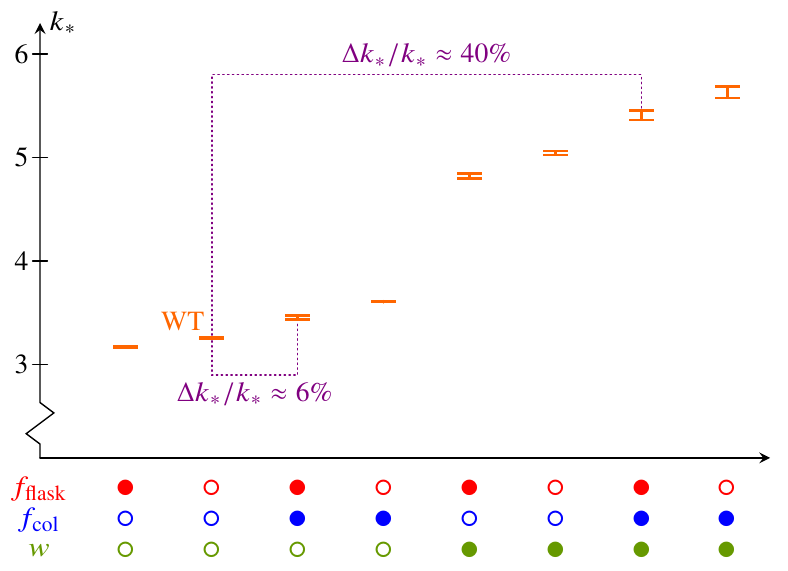}
\caption{Effect of inhibiting contraction. Critical curvature $k_\ast$ against parameter sets: open circles indicate values identical to the estimated `wild-type' values (WT); filled circles indicate modified parameter values. Modified parameter values corresponding to the chemical treatments of Ref.~\cite{nishii99} are: $f_{\text{flask}}=f_{\text{col}}=1$, $w=0.37$.}
\label{fig:cont}
\end{figure}

We are left to discuss the observations of Nishii and Ogihara~\cite{nishii99}, who showed that inversion of the \emph{Volvox carteri} embryo is arrested if actomyosin-mediated contraction is inhibited by various chemical treatments. They argued that it is the resulting lack of contraction of the spindle-shaped cells in the posterior, i.e. the relative expansion of the inverted part of the cell sheet, where the cells are columnar (Fig.~\ref{fig:volvox}b), that arrests inversion, the posterior hemisphere being swollen compared to the inverted part of the cell sheet. We therefore model actomyosin inhibition by setting $f_{\text{flask}}=f_{\text{col}}=1$. This modification does not however increase the critical curvature $k_\ast$ very much (Fig.~\ref{fig:cont}). Accordingly, if the arrest of inversion of the treated embryos were solely caused by this lack of relative expansion, it would follow that inversion operates quite close to its mechanical limit.

There is, however, a curious observation that appears, almost as a footnote, in the caption of Fig.~6 of Ref.~\cite{nishii99}: in embryos in which contraction had been inhibited, the number of cells constituting the bend region was smaller than in untreated embryos. While it is unclear why the chemical treatments applied in Ref.~\cite{nishii99} should have this effect, it can be introduced into the model by reducing the value of $w$. To estimate the magnitude of this effect very roughly, we turn to previously published experimental figures: in untreated embryos, about 7 cells make up the bend region (Fig.~2c--e in Ref.~\cite{viamontes77}); in treated embryos, this is reduced to about 4 (Fig.~6d,e in Ref.~\cite{nishii99}). We therefore estimate $w\approx 0.37$ in the treated embryos. With this value of the width of the bend region, the critical curvature $k_\ast$ is increased considerably (Fig.~\ref{fig:cont}), suggesting that inversion does not need to be close to its mechanical limit to explain the observed arrest of inversion.

What is more, cells in the bend region of the treated embryos are less markedly flask-shaped than those in the untreated ones (Fig.~6 in Ref.~\cite{nishii99}), which might indicate that the intrinsic curvature $\kappa_{\text{flask}}$ is reduced in the untreated embryos. This reduction of the intrinsic curvature may provide another explanation for the failure of actomyosin-inhibited embryos to invert. The experimental images in Ref.~\cite{nishii99} suggest that, as in `inversion' of the InvA mutant~\cite{nishii03}, the lips start to peel back, but then fail to flip over completely, as in the shapes obtained in the model for low values of the intrinsic curvature~(Fig.~\ref{fig:bif}b).

Closer examination of the shapes of the treated embryos in Ref.~\cite{nishii99} suggests that the treated embryos are crammed into the embryonic vesicle that surrounds the embryos during inversion because of lack of contraction of the spindle-shaped cells. In fact, Ueki and Nishii~\cite{ueki09} studied the InvB mutant of \emph{Volvox carteri} in which the embryonic vesicle fails to grow properly during development. They showed that inversion is prevented in the InvB mutant by the confining forces of the embryonic vesicle: inversion completes if and only if the InvB mutant is microsurgically removed from the embryonic vesicle~\cite{ueki09}. Nishii and Ogihara~\cite{nishii99} reported that fragments of treated embryos removed from the embryonic vesicle can invert, but left open the question whether complete treated embryos can invert when removed from the embryonic vesicle. While the above discussion suggests that lack of relative expansion is not the mechanical reason for inversion failure in the treated embryo, this experiment could help to decide which of the three other candidate mechanisms is the dominant cause of inversion arrest: is it the reduction of the width of the bend region, the reduced intrinsic curvature in the bend region, or the confinement of a swollen embryo to the stiff embryonic vesicle that prevents the lips from flipping over completely? The final effect could also play a role in the InvA mutant since, as discussed previously, the maximal opening of the phialopore increases as the intrinsic curvature is reduced above the critical curvature (Fig.~\ref{fig:rphial}).

\subsection{Effective Energy}
\begin{figure}[b]
\centering\includegraphics{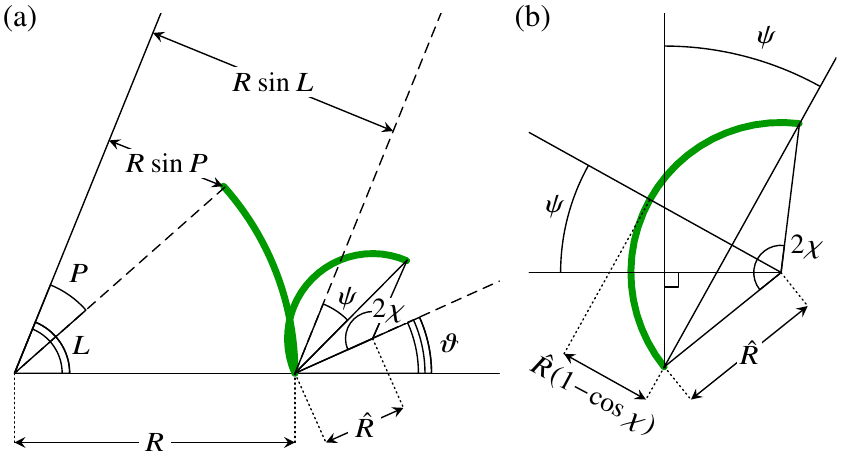}
\caption{Effective Geometry for Lips. (a)~Geometric considerations: a lip of length $\ell = R(L-P)$ on a shell of radius $R$ has folded into an arc of negative curvature and of radius $\hat{R}$ intercepting an angle~$2\chi$, rotated by an angle $\vartheta$ with respect to the original configuration. The chord intercepted by the lip makes an angle $\psi$ with the axis of the shell. (b)~Details for the calculation of the displacement of the midpoint of the lip.}
\label{fig:EEgeom}
\end{figure}

To gain some insight into the physical mechanism underlying the flipover of the lips, it is useful to consider a reduced (two-parameter) model that balances three physical effects:
\begin{enumerate}[leftmargin=*]
 \item the bending energy associated with deviations of the curvature of the bent lips from its intrinsic value;
 \item the stretching energy associated with the hoop stretches induced by the bending of the lips;
 \item the elastic energy of the formation of a second bend region that links the bent lips up to the remainder of the shell. 
\end{enumerate}
We begin by describing the reduced geometry: we consider an elastic spherical shell of undeformed radius $R$, with a phialopore of angular extent $P$ at its anterior pole (Fig.~\ref{fig:EEgeom}a). Cuts define $N$ lips of length $\ell=R(L-P)$ adjacent to this opening. As the shell deforms, these lips bend into circular arcs of radius $\hat{R}$ (and negative curvature), intercepting an angle~$2\chi$. Since stretching is energetically more costly than bending, there cannot be any stretching at leading order, and thus $2\chi\hat{R}=\ell$. In what follows, we non-dimensionalize lengths with $R$ and energy densities with $Eh$.

Further, the base of the lips may rotate by an angle $\vartheta$ with respect to the undeformed configuration; as a result, the chord intercepted by the lip makes an angle $\psi$ with the axis of the shell, where $\psi = L-\left(90^\circ-\chi+\vartheta\right)$. The distance from the tip of the deformed lip to the axis of the shell is thus
\begin{align}
r_{\mathrm{P}}=\sin{L}-2\hat{R}\sin{\chi}\cos{\left(L+\chi-\vartheta\right)}. 
\end{align}
Since we have already imposed that the meridional strains vanish globally, the no-stress condition at the free edge of the lips forces the hoop strains there to vanish at leading order. The azimuthal compression at the phialopore is thus
\begin{align}
\Phi_{\mathrm{P}}=\dfrac{\sin{P}}{\sin{L}-2\hat{R}\sin{\chi}\cos{\left(L+\chi-\vartheta\right)}}, 
\end{align}
and we let $\Psi_{\mathrm{P}}=\log{\Phi_{\mathrm{P}}}$. At the base of the lip, $\Phi_{\mathrm{B}}=1$ and thus $\Psi_{\mathrm{B}}=0$ to match up to the part of the shell without lips. We therefore approximate $\Psi'=-\log{\Phi_{\mathrm{P}}}/\ell$ to minimise the integral of ${\Psi'}^2$ along the lips. In particular, at the midpoint of the lip, $\Psi_{\mathrm{M}}=\tfrac{1}{2}\log{\Phi_{\mathrm{P}}}$, and thus $\Phi_{\mathrm{M}}=\sqrt{\Phi_{\mathrm{P}}}$. 

To describe the additional hoop strains resulting from the bending of the lips, we compute the distance of the midpoint of the deformed lip from the axis of revolution (Fig.~\ref{fig:EEgeom}b),
\begin{align}
r_{\mathrm{M}}=\sin{L}+\hat{R}\left[\sin{(L-\vartheta)}-\sin{(L+\chi-\vartheta)}\right].
\end{align}
The hoop strain is therefore $E_{\mathrm{M}}=\Phi_{\mathrm{M}}r_{\mathrm{M}}/\rho_{\mathrm{M}}-1$ at the midpoint of the lips, with $\rho_{\mathrm{M}}=\sin{\left(P+\ell/2\right)}$. 

Let $R_0$ denote the intrinsic radius of curvature of the lips, and let $\varepsilon\ll1$ be the non-dimensional thickness of the shell. The effective elastic energy is then $\mathcal{F}=\mathcal{F}_1+\mathcal{F}_2+\mathcal{F}_3$, the sum of the three respective contributions of the physical effects described above:
\begin{align}
\mathcal{F}_1&=\rho_{\mathrm{M}}\ell\,\varepsilon^2\left(\dfrac{1}{\hat{R}}-\dfrac{1}{R_0}\right)^2,&\mathcal{F}_2&=\rho_{\mathrm{M}}\ell\, E_{\mathrm{M}}^2,&\mathcal{F}_3&=\varepsilon^{3/2}\vartheta^2.
\end{align}
In these expressions, $\varepsilon$ is the non-dimensional bending modulus, and the factor $\rho_{\mathrm{M}}\ell$ corresponds to integration over the lips. The scaling of the prefactor of the final term is inspired by the energetics of a Pogorelov dimple \cite{landaulifshitz}. As announced, this effective energy has reduced the number of parameters in the problem to two, viz. the radius $\hat{R}$ of the deformed lip and the angle $\vartheta$.

\begin{figure}
\centering\includegraphics{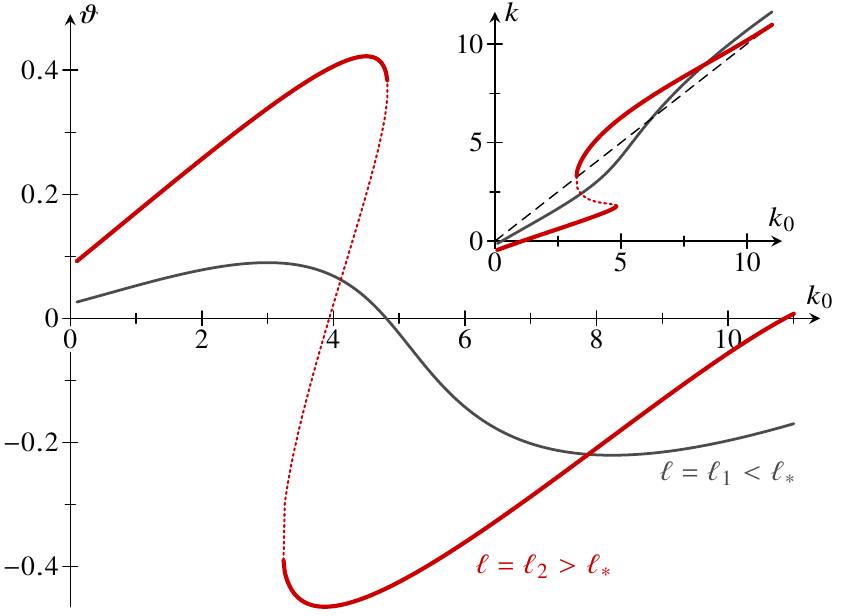}
\caption{Effective Energy of Lips. Coordinates of the minimum of the effective energy $\mathcal{F}$ in the $(k_0,\vartheta)$ diagram for parameter values $\ell_1=0.37<\ell_\ast$, $\ell_2=0.55>\ell_\ast$, illustrating flip-over of lips at large enough values of $k_0$. Folds arise for $\ell>\ell_\ast$; dotted lines mark positions of unstable saddle points. Inset: corresponding plot in the $(k_0,k)$ diagram.}
\label{fig:EEres}
\end{figure}

We determine minima of $\mathcal{F}$ numerically using \textsc{Mathematica} (Wolfram, Inc.). Denoting by $k_0=1/R_0$ and $k=1/\hat{R}$ the intrinsic and actual curvatures of the bend region, we plot the position of the energy minima in the $(k_0,\vartheta)$ and $(k_0,k)$ diagrams (Fig.~\ref{fig:EEres}), for different values of the lip extent $\ell$. At large enough values of $k_0$, $k>k_0$ and $\vartheta<0$, so the lips bend and rotate outwards and thus flip over. At small values of $k_0$, $\vartheta>0$ and $k<k_0$: the lips resist bending and flipping over by rotating inwards to alleviate hoop strains. We also note that a critical value $\ell_\ast$ separates two kinds of behavior: if $\ell<\ell_\ast$, the transition between the two states is continuous, but becomes discontinuous if $\ell>\ell_\ast$, with the two states coexisting in an intermediate range of $k_0$. Given the existence of other solution branches discussed previously, this behavior is not surprising; the discontinuous transition signals a break-down of the geometric approximation of uniformly curved lips as the lip extent grows. 

Nonetheless, this discussion shows how the behavior observed in the continuum model can be attributed to three simple physical effects. Conversely, if any of these three effects is not considered, the reduced model fails to capture the observed behavior: clearly, if $\mathcal{F}_1$ is neglected, there is no dependence on $R_0$, and if $\mathcal{F}_2$ is neglected, there is no coupling between $\hat{R}$ and $\vartheta$, and there is a minimum $\hat{R}=R_0$, $\vartheta=0$ for all $R_0$. Finally, if the contribution of $\mathcal{F}_3$ is not considered, we find, numerically, two minima with $\hat{R}=R_0$ and $\vartheta\not=0$ for all $R_0$, so that the transition between the two kinds of behavior discussed above is not reproduced. Hence all three of these physical effects are essential to explain the observed behavior.

\section{Conclusion}
In this paper, we have derived a simple, averaged elastic theory to analyze the flip-over of lips observed during type-A inversion in \emph{Volvox carteri}, and showed by means of a reduced model how the observed behavior can be attributed to three geometrical effects. The model can explain the observations of Nishii, \emph{et al.} on the InvA mutant~\cite{nishii03}, but suggests that the failure of embryos treated with actomyosin inhibitors to invert~\cite{nishii99} does not result from lack of relative expansion of the cell sheet. Several potential candidate mechanisms for the inversion arrest in the treated embryos, remain however. Further experiments on embryos removed from the embryonic vesicle could help to distinguish between these three mechanisms discussed above and in particular clarify the role of the confinement to the embryonic vesicle for both the chemically treated embryos and those of the InvA mutant.

Further experiments could also allow taking the present, qualitative analysis to a more quantitative level. The large size of the gonidia of \emph{Volvox carteri} at the inversion stage may hamper such a comparison between theory and experiment. This difficulty could be addressed by using the gonidialess mutants of \emph{Volvox carteri} that invert normally~\cite{tam91} or other type-A inverters such as \emph{Volvox gigas} that have smaller gonidia at the inversion stage~\cite{pocock33}.

It also remains unclear whether $N=4$ lips are mechanically optimal for inversion in some sense. In the much more dynamic process of impact petalling of metallic sheets~\cite{wierzbicki99}, during which lips similar to those seen in inversion arise, the number of lips is set by minimizing dissipation of elastic and plastic energy. It is tempting to argue that a similar mechanical trade-off arises for inversion: the more lips there are, the easier they are to invert, but the coupling of the lips to the remainder of the shell and hence the extent to which the lips aid inversion of the connected part of the shell decrease with increasing lip number. While this mechanical argument does suggest an intermediate optimal lip number, it ignores the rather more combinatorial constraint that cells must divide in a way that robustly defines the lips. A very recent study~\cite{imranalsous18} of a related cell packing problem in \emph{Drosophila} egg chambers highlighted the role of entropic effects in selecting the spatial structure of connected and unconnected adjacent cells. In \emph{Volvox carteri}, the four lips are defined around the 16-cell stage~\cite{green81a}. It is tempting to speculate that this number of lips is stabilized by similar entropic effects, especially since a larger lip number could only be defined at later stages of cell division, with a combinatorial explosion of possible packings. The cell division pattern of \emph{Volvox carteri} has been mapped up to and including the 64-cell stage~\cite{green81a}. Extending this work to later cell division stages could shed more light on these issues.

All the algae of the family Volvocaceae display some kind of inversion~\cite{matt16,hallmann06}, although, interestingly, the genus \emph{Astrephomene} of the closely related family Goniaceae forms spherical colonies of up to 128 cells without the need for inversion~\cite{yamashita16}. Our analysis should therefore finally be considered in the context of the evolution of Volvocaceae. The present analysis of type-A inversion in \emph{Volvox carteri} indicates that relative expansion of different parts of the cell sheet is not mechanically required for this inversion (although it may play a role in enabling inversion within the confinement of the embryonic vesicle). By contrast, we have previously shown that the peeling of the anterior hemisphere (Fig.~\ref{fig:volvox}c) during type-B inversion in \emph{Volvox globator} is mainly driven by contraction of parts of the cell sheet~\cite{haas18}. This contraction is regulated separately from the earlier invagination of the cell sheet, and therefore indicative of a transition towards higher developmental complexity within Volvocaceae~\cite{haas18}. The present result that type-A inversion does not rely on this additional deformation mode lends further support to this inference. This additional complexity in type-B inversion thus appears as the geometric price of the absence of lips. The questions how these different features --- formation of lips and separately regulated inversion subprocesses --- evolved, and in particular, whether they were lost from an ancestral alga, remain widely open. The recent observation that inversion in the genus \emph{Pleodorina} features non-uniform cell shape changes~\cite{hohn16}, shared with type-B inversion~\cite{hohn11} but absent from type-A inversion, might begin to shed some light on these issues.

\begin{acknowledgments}
We thank Stephanie H\"ohn for many useful discussions about the biology of inversion and comments on a draft of this paper, and are grateful for support from 
the Engineering and Physical Sciences Research Council (Established Career Fellowship EP/M017982/1, REG; Doctoral Prize Fellowship, PAH), the Schlumberger Chair Fund, the Wellcome Trust 
(Investigator Award 207510/Z/17/Z, REG) and Magdalene College, Cambridge (PAH).
\end{acknowledgments}

\appendix \section{Geometric Simplifications}\label{appB}
In this appendix, we discuss the geometric approximations in more detail. For a general geometric description of the lips, we must replace Eq.~(\ref{eq:surf}) with 
\begin{align}
\vec{r}(s,\phi)=\overline{r}(s,\phi)\,\vec{u_r}\Bigl(\overline{\phi}(s,\phi)\Bigr)+\overline{z}(s,\phi)\,\vec{u_z}.
\end{align}
By symmetry, $\overline{\phi}$ is an odd function of $\phi$, while $\overline{r}$ and $\overline{z}$ are even functions of $\phi$. Expanding,
\begin{subequations}
\begin{align}
\overline{r}(s,\phi)&=r(s)+R(s)\phi^2+\mathcal{O}\bigl(\phi^4\bigr),\\
\overline{\phi}(s,\phi)&=\Phi(s)\phi+\mathcal{O}\bigl(\phi^3\bigr),\\
\overline{z}(s,\phi)&=z(s)+Z(s)\phi^2+\mathcal{O}\bigl(\phi^4\bigr).
\end{align}
\end{subequations}
In particular, merely on grounds of symmetry, we thus expect variations of $\overline{\phi}$ across the lips to be of order $\mathcal{O}(\varphi)=\mathcal{O}(N^{-1})$. Hence, at least in the limit of a large number of lips, these variations swamp those of $\overline{r}$ and $\overline{z}$, which result from $R$ and $Z$ and are thus expected to be of order $\mathcal{O}(\varphi^2)=\mathcal{O}(N^{-2})$. The argument thus suggests that the largest deformations are those of simple azimuthal compression, $\overline{\phi}(s,\phi)=\Phi(s)\phi$, to which we have restriced at the start of our analysis. 

It is however important to note that this does \emph{not} imply that the elastic theory we have derived is the asymptotic limit of the general theory in the limit of a large number of lips. It is not hard to see that in fact, a theory that is asymptotically exact to order $\mathcal{O}(\varphi^2)$ depends on the corrections to $\overline{r}$ and $\overline{z}$ up to order $\mathcal{O}(\varphi^4)$. Despite the geometric restriction that the simple azimuthal compression that we have imposed in Eq.~(\ref{eq:phibar}) therefore implies, we note that the theory that results from it and that we have analysed in this paper is simple enough to allow some detailed analysis, yet features a crucial coupling between the meridional and circumferential deformations resulting from $\Phi$. The importance of this coupling is revealed by the observation that, in its absence, the lips can zero their contribution to the energy density by adopting their intrinsic meridional stretches and curvatures (as one-dimensional elastic filaments would do), and then compressing azimuthally at no energetic cost to make the circumferential strain and curvature strain vanish. This decouples the lips from the remainder the shell, so their flipping over does not help the remainder of the shell to invert.  

\section{Governing Equations}\label{appA} 
In this appendix, we derive the governing equations associated with the elastic energy (\ref{eq:E}). We note the variations
\begin{subequations}
\begin{align}
\delta E_s&=\dfrac{1}{f_s^0}\Bigl(\sec{\beta}\,\delta r'+f_s\tan{\beta}\,\delta\beta\Bigr),\\
\delta E_\phi&=\dfrac{1}{f_\phi^0}\left(f_\phi\Phi\,\delta\Psi+\dfrac{\Phi}{\rho}\,\delta r\right)
\end{align}
and
\begin{align}
\delta K_s&=\dfrac{\delta\beta'}{f_s^0},\\\delta K_\phi&=\dfrac{1}{f_\phi^0}\left(f_\phi\kappa_\phi\Phi\,\delta\Psi+\Phi\dfrac{\cos{\beta}}{\rho}\delta\beta\right),
\end{align}
\end{subequations}
which are obtained from the definitions of the strains. The variation of the elastic energy takes the form
\begin{align}
\dfrac{\delta\mathcal{E}}{2\pi\rho}=n_s\delta E_s+n_\phi\delta E_\phi+m_s\delta K_\phi+m_\phi\delta K_\phi+\dfrac{\Xi}{\rho}\delta\Psi',
\end{align}
wherein the shell stresses are
\begin{widetext}
\begin{subequations}\label{eq:n6}
\begin{align}
n_s&=\dfrac{Eh}{1-\nu^2}\biggl[E_s+\nu E_\phi+a\lambda^2{\Psi'}^2\biggr]+\dfrac{Eh^3}{12(1-\nu^2)}\biggl[3{\kappa_s^0}^2E_s+\nu\Bigl({\kappa_s^0}^2+\kappa_s^0\kappa_\phi^0+{\kappa_\phi^0}^2\Bigr)E_\phi-2\kappa_s^0K_s\nonumber\\
&\hspace{80mm}-\nu\bigl(\kappa_s^0+\kappa_\phi^0\bigr)K_\phi+a\lambda^2{\Psi'}^2\Bigl(6{\kappa_s^0}^2-6\kappa_s^0\kappa_\phi^0+{\kappa_\phi^0}^2\Bigr)\biggr],\label{eq:ns}\\
n_\phi&=\dfrac{Eh}{1-\nu^2}\biggl[E_\phi+\nu E_s+\nu a\lambda^2{\Psi'}^2\biggr]+\dfrac{Eh^3}{12(1-\nu^2)}\biggl[3{\kappa_\phi^0}^2E_\phi+\nu\Bigl({\kappa_s^0}^2+\kappa_s^0\kappa_\phi^0+{\kappa_\phi^0}^2\Bigr)E_s-2\kappa_\phi^0K_\phi\nonumber\\
&\hspace{93mm}-\nu\bigl(\kappa_s^0+\kappa_\phi^0\bigr)K_s+a\lambda^2{\Psi'}^2\kappa_s^0\bigl(3\kappa_s^0-2\kappa_\phi^0)\biggr],
\end{align}
the shell moments are
\begin{align}
m_s&=\dfrac{Eh^3}{12(1-\nu^2)}\biggl[K_s+\nu K_\phi-2\kappa_s^0 E_s-\nu\bigl(\kappa_s^0+\kappa_\phi^0\bigr)E_\phi-a\lambda^2{\Psi'}^2\bigl(3\kappa_s^0-2\kappa_\phi^0\bigr)\biggr],\label{eq:ms}\\
m_\phi&=\dfrac{Eh^3}{12(1-\nu^2)}\biggl[K_\phi+\nu K_s-2\kappa_\phi^0 E_\phi-\nu\bigl(\kappa_s^0+\kappa_\phi^0\bigr)E_s-\nu a\lambda^2{\Psi'}^2\bigl(2\kappa_s^0-\kappa_\phi^0\bigr)\biggr],
\end{align}
and where
\begin{align}
\Xi&=2a\rho\lambda^2\Psi'\left\{\vphantom{\left(\dfrac{A^2}{A^2}\right)^2}\dfrac{Eh}{1-\nu^2}\left[E_s+\nu E_\phi+\dfrac{1-\nu}{2}\right]-\dfrac{Eh^3}{12(1-\nu^2)}\Biggl[\bigl(3\kappa_s^0-2\kappa_\phi^0\bigr)K_s+\nu(2\kappa_s^0-\kappa_\phi^0\bigr)K_\phi\right.\nonumber\\
&\hspace{30mm}-\left.\Bigl(6{\kappa_s^0}^2-6\kappa_s^0\kappa_\phi^0+{\kappa_\phi^0}^2\Bigr)E_s-\nu\kappa_s^0\bigl(3\kappa_s^0-2\kappa_\phi^0\bigr)E_\phi-\dfrac{1-\nu}{2}\Bigl(3{\kappa_s^0}^2-4\kappa_s^0\kappa_\phi^0+{\kappa_\phi^0}^2\Bigr)\Biggr]\vphantom{\left(\dfrac{A^2}{A^2}\right)^2}\right\}\nonumber\\
&\hspace{10mm}+\dfrac{18}{5}a^2\rho\lambda^4{\Psi'}^3\left\{\vphantom{\left(\dfrac{A^2}{A^2}\right)^2}\dfrac{Eh}{1-\nu^2}+\dfrac{Eh^3}{12(1-\nu^2)}\Bigl(10{\kappa_s^0}^2-16\kappa_s^0\kappa_\phi^0+6{\kappa_\phi^0}^2\Bigr)\right\}.\label{eq:Xi}
\end{align}
\end{subequations}
Finally, upon letting
\begin{align}
&N_s=\dfrac{n_s}{f_\phi f_s^0},&& N_\phi=\Phi\dfrac{n_\phi}{f_sf_\phi^0},&&M_s=\dfrac{m_s}{f_\phi f_s^0},&& M_\phi=\Phi\dfrac{m_\phi}{f_sf_\phi^0},
\end{align}
the variation becomes
\begin{align}
\dfrac{\delta\mathcal{E}}{2\pi}&=\Bigl\llbracket r N_s\sec{\beta}\,\delta r+r M_s\,\delta\beta+\Xi\,\delta\Psi\Bigr\rrbracket-\int{\Biggl\{\dfrac{\mathrm{d}}{\mathrm{d}s}\Bigl(r N_s\sec{\beta}\Bigr)-f_sN_\phi\Biggr\}\delta r\,\mathrm{d}s}\nonumber\\
&\hspace{15mm}+\int{\left\{\dfrac{\mathrm{d}\Xi}{\mathrm{d}s}-r f_s\bigl(N_\phi+\kappa_\phi M_\phi\bigr)\right\}\delta\Psi\,\mathrm{d}s}+\int{\Biggl\{\dfrac{\mathrm{d}}{\mathrm{d}s}\Bigl(r M_s\Bigr)-r f_sN_s\tan{\beta}-M_\phi\cos{\beta}\Biggr\}\delta\beta\,\mathrm{d}s}. \label{eq:var}
\end{align}
As in standard shell theories~\cite{libai}, we define the transverse shear tension $T=-N_s\tan{\beta}$ to remove a singularity in the resulting equations, so that
\begin{subequations}\label{eq:goveq}
\begin{align}
\dfrac{\mathrm{d}N_s}{\mathrm{d}s}&=f_s\left(\dfrac{N_\phi-N_s}{r}\cos{\beta}+\kappa_s T\right),\label{eq:goveq1}&
\dfrac{\mathrm{d}M_s}{\mathrm{d}s}&=f_s\left(\dfrac{M_\phi-M_s}{r}\cos{\beta}-T\right),&
\dfrac{\mathrm{d}\Xi}{\mathrm{d}s}&=r f_s\bigl(N_\phi+\kappa_\phi M_\phi\bigr).
\end{align}
By differentiating the definition of $T$ and using the first of (\ref{eq:goveq1}), one finds that
\begin{align}
\dfrac{\mathrm{d}T}{\mathrm{d}s}=-f_s\left(\kappa_sN_s+\kappa_\phi N_\phi+\dfrac{T}{r}\cos{\beta}\right). 
\end{align}
\end{subequations}
\end{widetext}
Together with the geometrical equations $r'=f_s\cos{\beta}$ and $\beta'=f_s\kappa_s$, equations (\ref{eq:goveq}) describe the deformed shell. The final shape equation $z'=f_s\sin{\beta}$ is redundant. From the boundary terms in Eq.~(\ref{eq:var}), we deduce the seven pertaining boundary conditions:
\begin{subequations}
\begin{align}
r&=0, &\beta&=0, &T&=0,&\Phi&=1\Longrightarrow\Psi=0,\\
&&&&&&&\makebox[0pt][l]{at the posterior,}\nonumber\\
N_s&=0,&M_s&=0,&\Xi&=0,&&\\
&&&&&&&\makebox[0pt][l]{at the phialopore.}\nonumber
\end{align}
\end{subequations}
\subsubsection*{Remark on the numerical solution of Eqs.~(\ref{eq:goveq})}
At each stage of the numerical solution, $f_\phi$ and $\kappa_\phi$ are computed directly from $r$ and $\beta$, but it is less straightforward to compute $\Psi'$, $E_s$, $K_s$ from $\Xi$, $N_s$, $M_s$, required in order to obtain $f_s$, $\kappa_s$, $N_\phi$, $M_\phi$ and hence continue the integration. To this end, we eliminate $E_s$, $K_s$ between  Eqs.~(\ref{eq:n6}a,c,e) by solving a linear system of equations to obtain a cubic equation for~$\Psi'$, which is solved exactly using the algorithm described in Ref.~\cite{nrecipes}. Once $\Psi'$ is known, Eqs.~(\ref{eq:n6}a,c) become a linear system of equations for $E_s$, $K_s$. From these, $f_s$, $\kappa_s$, $N_\phi$, and $M_\phi$ can be computed and the integration can be continued.

\bibliography{inv3}
\end{document}